\documentclass[12pt,a4paper]{article}
\newcommand{\be}{\begin{equation}}
\newcommand{\en}{\end{equation}}
\newcommand{\bea}{\begin{eqnarray}}
\newcommand{\ena}{\end{eqnarray}}

\oddsidemargin .7cm \textwidth 14.5cm \textheight 23cm  \voffset -1cm \topmargin 0 cm \fontsize{12pt}{14pt}\selectfont
\begin{document}

\title{\textbf{Energy distribution of 2+1 dimensional black holes with nonlinear electrodynamics}}

\author{Leonardo Balart \\    \\ \small CEFIMAS and  Departamento de F\'{\i}sica, \\  \small Universidad Nacional de La Plata,\\
\small C.C. 67, (1900), La Plata, Argentina\\ \\
\emph{ \small lbalart@fisica.unlp.edu.ar}}

\date{}

\maketitle

\begin{abstract}
The energy distributions for a black hole solution resulting from coupling electrodynamics and gravity in 2+1 dimensions are obtained. This solution considers the correction for a 2+1 static charged black hole from the first contribution of the weak field limit of one loop QED in 2+1 dimensions. The Einstein and M{\o}ller energy-momentum prescriptions are used to evaluate the energy distributions associated with the mentioned 2+1 dimensional black hole and other 2+1 black hole solutions coupled with nonlinear electrodynamics. A relation that connects the coefficients of both prescriptions is established.
\end{abstract}



\vspace{.4cm}
\section{Introduction}

 Due to general covariance of the theory, there is not a single way to define an energy-momentum tensor in the context of general relativity. Thus, since the introduction of the Einstein energy-momentum complex (or pseudotensor)~\cite{einstein}, many other energy-momentum complexes have been introduced~\cite{tolman,papapetrou,landau,bergman,weinberg,misner,qadir}; all of them restricted to evaluate the energy distribution in quasi-Cartesian coordinates. M{\o}ller~\cite{moller} proposed an expression for an energy-momentum complex which could be used to calculate the energy distribution inside a spherical surface for the gravitational field in any coordinates system and not only in quasi-Cartesian coordinates. In the recent years, numerous researches have been accomplished on evaluating the energy distributions of different space-time solutions using the energy-momentum complexes (see for example Refs.~\cite{Vagenas:2006pj} and \cite{Matyjasek:2008ku} and references cited therein).

Some authors have studied the energy localization in 2+1 dimensional models using energy-momentum complexes. Virbhadra obtained the energy-momentum complexes in quasi-Cartesian coordinates for two exact solutions of the Einstein equation~\cite{virbhadra95}. Yang and Radinschi calculated energy distributions of different black hole solutions using Einstein and M{\o}ller energy-momentum complexes~\cite{Yang:2003vh} and also the Landau-Lifshitz and Weinberg energy-momentum complexes~\cite{Radinschi:2007tt}. The M{\o}ller energy-momentum complex was also used to obtain the energy distribution of a radial magnetic field in AdS~\cite{Grammenos:2004ys} and of an Einstein-Klein-Gordon system~\cite{Radinschi:2005cj} respectively. Vagenas evaluated energy distributions for non-static spinless~\cite{Vagenas:2003fc} and rotating BTZ black holes~\cite{Vagenas:2004zt}. Besides other formulations have been applied to calculate the gravitational energy of black holes in 2+1 dimensions: the Brown-York method~\cite{brown} (see also Refs.~\cite{Balasubramanian:1999re} and \cite{Miskovic:2006tm}) and the gravitational teleparallelism~\cite{Sousa:2003sx}.

In this paper we investigate the static charged BTZ metric~\cite{btz} with nonlinear electromagnetic field which arises from the correction of the first contribution of the weak field limit of one loop QED in 2+1 dimensions.
We use energy-momentum prescriptions of Einstein and M{\o}ller to investigate the energy distributions of the obtained solution in this work. It is also interesting to consider other 2+1 dimensional electrodynamic theories of nonlinear type coupled to the static charged 2+1 black hole with cosmological constant given in Refs.~\cite{cat.gar} and \cite{cat.gar2} to get the corresponding energy distributions.
Thus, it is shown there is a relation that connects the ``coefficients" of the expression of the Einstein energy distribution which is of the form $E(r) = \sum \alpha_n^{(E)}r^{-n}$ with those of the expression of M{\o}ller energy distribution, i.e. $E(r) = \sum \alpha_n^{(M)}r^{-n}$, where in both cases we can define $\alpha_0^{(E)}$ and $\alpha_0^{(M)}$ as functions depending of $r$. We have to mention that Vagenas obtained the analog connection in 3+1 dimensions (see Ref.~\cite{Vagenas:2006pj}). However, as was shown by Vagenas in Ref.~\cite{Vagenas:2003ug}, such a relation does not hold in 1+1 dimensions since he found that the Einstein energy-momentum complex is zero, thus it is not possible to establish a relation.

The paper is organized as follows. In Section 2 we calculate for the static BTZ solution considering QED in 2+1 dimensions, the counterpart of the correction for the Reissner-Nordstr\"{o}m metric from the first contribution of the weak field limit of one loop QED in 3+1 dimensions. In the next two sections we find the energy distributions of the obtained metric. Sections 3 and 4 show the results obtained for the energy distribution using the Einstein and M{\o}ller prescriptions, respectively. The same calculations were performed for other kinds of static black hole solutions in 2+1 dimensions coupled with nonlinear electrodynamics. The results are showed in Section 5. Finally, in Section 6  we summarize the results.

\vspace{.4cm}
\section{The minimal coupling between gravitation and nonlinear electrodynamics in 2+1 dimensions}

In this section we perform the coupling between electrodynamics and gravity in 2+1 dimensions. We consider
the correction for the static charged black hole with cosmological constant from the first contribution of the weak field limit of one loop QED in 2+1 dimensions.

The Einstein theory coupled with the nonlinear electrodynamics in 2+1 dimensions
arises from the following action
\be
S = \int\sqrt{-g}\left[\frac{1}{2 \kappa}(R - 2\Lambda) + L(F)\right]d^3x
\,\,\label{action}, \en
where $\kappa = 8 \pi$, $g$ is the determinant of the metric tensor $g_{\mu\nu}$, $\Lambda$ is the cosmological constant and $L(F)$ is an electromagnetic lagrangian depending on $F \equiv F^{\mu\nu}F_{\mu\nu}$.
Varying this action with respect to gravitational field we get the Einstein equations
\be
G_{\mu\nu} + \Lambda g_{\mu\nu} = \kappa T_{\mu\nu}
\,\,\label{einstein}, \en
where the corresponding stress-energy tensor is given by
\be
T_{\mu\nu} = g_{\mu\nu}L(F) - 4 \, L,_F F_{\mu\alpha}F_\nu^{\,\,\alpha}
\,\,\label{e-m} \en
and $L,_F$ denotes the derivative of $L(F)$ with respect to $F$.
The electromagnetic field equations are
\be
\nabla_\mu(F^{\mu\nu}L,_F)= 0
\,\,\label{e-m1}. \en
For the static and spherically symmetric 2+1 dimensional solution, the metric is given by
\be
ds^2 = -A(r)dt^2 + \frac{1}{A(r)}dr^2 + r^2 d\theta^2
\,\,\label{metric}, \en
where $A(r)$ is a metric function to be determined.

Considering only the non zero component, the electromagnetic tensor becomes $F^{tr} = f(r)$.
Then, Eqs.~(\ref{einstein}-\ref{e-m1}) implies
\be
\frac{1}{2 r} \frac{\partial A(r)}{\partial r} = \kappa [L(F) + 4 f^2(r) L,_F] - \Lambda
\,\,\label{equation} \en
\be
f(r)L,_F  = -\frac{Q}{8 \pi r}
\,\,\label{charge}. \en

From the analysis of the weak field limit of the complete one-loop
approximation of QED in 2+1 dimensions (see Ref.~\cite{Ritz:1995nt},
where it was analyzed the case of four photon scattering in any dimension $d\geq 2$),
the effective lagrangian is given by
\be
L = -\frac{1}{4 \pi}F + \frac{\mu}{4 \pi} F^2
\,\,\label{hei-euler2}, \en
where $\mu = e^4/(480 \, \pi^2 m_e^5)$. The invariant $F = -2 f(r)^2$ and Eq. (\ref{equation}),
allows us to write
\be
A(r) = - M - \Lambda r^2 - 2 \kappa \int  \, \left[\frac{r}{2 \pi} f^2(r) + \frac{3}{\pi} r \mu f^4(r)\right] dr
\,\,\label{int-fn}. \en
Moreover, from Eqs.~(\ref{charge}) and (\ref{hei-euler2}) result
\be
f(r) = \frac{Q}{2 r} - \frac{\mu Q^3}{2 r^3}
\,\,\label{efe}, \en
which permits to obtain an explicit expression for the metric function
\be
A(r) = -M - \Lambda r^2 - 2 Q^2\ln r - \frac{\mu Q^4}{2 r^2} + O(\mu^2)
\,\,\label{fn}. \en
Note that for the limit case $\mu = 0$ in Eq.~(\ref{fn}) the static charged BTZ solution is obtained.
The case of QED to one loop in 3+1 dimensions for the Reissner-Nordstr\"{o}m metric has been considered in Ref.~\cite{DeLorenci:2001bd}.

In the next two sections we use the Einstein and M{\o}ller prescriptions to calculate the energy distributions of the modified static charged BTZ metric.

\vspace{.4cm}
\section{Energy distribution under the Einstein \\ prescription}

In general relativity the conservation law must be valid for all frames of reference, thus if $T^{\,\,\, \nu}_\mu$ is the symmetric energy-momentum tensor of matter, it satisfies
\be
{T^{\,\,\, \nu}_\mu}_{;\nu} = 0
\,\,\label{conserv} \, , \en
The quantity that satisfies the conservation law in the usual sense is called energy-momentum complex which is given by
\be
\Theta^{\,\,\, \nu}_\mu = \sqrt{-g} \, (T^{\,\,\, \nu}_\mu + t^{\,\,\, \nu}_\mu)
\,\,\label{complex} \, , \en
where $t^{\,\,\, \nu}_\mu$ is an energy-momentum pseudotensor for the gravitational field. This complex can be written as the divergence of an antisymmetric superpotential~\cite{vonfreud}
\be
\Theta^{\,\,\, \nu}_\mu = \frac{1}{2\kappa}{H_\mu^{\nu\lambda}}_{,\lambda}
\,\,\label{Theta} \, , \en
where the superpotential is of the form
\be
H_\mu^{\nu\lambda} = \frac{g_{\mu\sigma}}{\sqrt{-g}}
[-g (g^{\nu\sigma} g^{\lambda\rho} - g^{\lambda\sigma} g^{\nu\rho})]_{,\rho}
\,\,\label{Hache} \, . \en
The Einstein energy-momentum complex satisfies the local conservation equation
\be
{\Theta^{\,\,\, \nu}_\mu }_{,\nu}= 0
\,\,\label{Cons.theta} \, . \en
The energy and the momentum components in quasi-Cartesian coordinates in 2+1
dimensions are given by
\be
P_\mu = \frac{1}{2\kappa} \int H_\mu^{0 i} \, n_i dl
\,\,\label{Ener-Mom} \, , \en
where $i = 1, 2$ and $n_i$ is the normal vector to the closed line $l$, and using the Gauss theorem we obtain the energy component
\be
E_E(r) = \frac{1}{2\kappa} \int {H_0^{0 i}} \, n_i dl
\,\,\label{Ener} \, . \en
In this prescription, the  nonzero components $H_0^{0 i}$ of the Einstein energy-momentum
complex are
\be
H_0^{0 1} = \frac{x}{r^2} (- M - \Lambda r^2 - 2 Q^2 \ln r - \mu \frac{Q^4}{2 r^2} -1)
\,\,\label{hache1}\, , \en
\be
H_0^{0 2} = \frac{y}{r^2} (- M - \Lambda r^2 - 2 Q^2 \ln r - \mu \frac{Q^4}{2 r^2} -1)
\,\,\label{hache2}\, . \en
Then, the energy distribution inside a circle with radius $r$ becomes
\be
E_E(r) = \frac{\pi}{\kappa}(- M - \Lambda r^2 - 2 Q^2 \ln r - \mu \frac{Q^4}{2 r^2} -1)
\,\,\label{Energy} \, . \en

\vspace{.4cm}
\section{Energy distribution under the M{\o}ller \\ prescription}

The M{\o}ller energy-momentum complex differs from the one corresponding to the Einstein prescription by an added quantity $S^{\,\,\, \nu}_\mu$
which verify ${S^{\,\,\, \nu}_\mu}_{,\nu}=0$. The definition of M{\o}ller energy-momentum complex is given by~\cite{moller}
\be
M^{\,\,\, \nu}_\mu = \frac{1}{\kappa}{\chi_\mu^{\nu\lambda}}_{,\lambda}
\,\,\label{Mol} \, , \en
where $\chi_\mu^{\nu\lambda}$ is an antisymmetric superpotential of the form
\be
\chi_\mu^{\nu\lambda} = \sqrt{-g} \,({g_{\mu\sigma}}_{,\rho}-
{g_{\mu\rho}}_{,\sigma})
g^{\nu\rho} g^{\lambda\sigma}
\,\,\label{Chi} \, . \en
The local conservation equation has the form
\be
{M^{\,\,\, \nu}_\mu }_{,\nu}= 0
\,\,\label{Cons.Mol} \, . \en
The energy and momentum components calculated using the M{\o}ller energy-momentum complex
are
\be
P_\mu = \frac{1}{\kappa} \int {\chi_\mu^{0 i}}_{, i} \, d^2x
\,\,\label{Ener-Mom.M} \, , \en
where $i = 1, 2$. Through the Gauss theorem the corresponding M{\o}ller energy result
\be
E_M(r) = \frac{1}{\kappa} \int {\chi_0^{0 i}}_{, i} \, dr d\theta
\,\,\label{Ener.M} \, . \en
At last, the nonzero component of the superpotential is
\be
\chi_0^{0 1} = 2\Lambda r^2 + 2 Q^2  - \mu \frac{Q^4}{r^2}
\,\,\label{chi1}\, . \en
Solving the integral in (\ref{Ener.M}) considering (\ref{chi1}) the energy distribution in
a circular region of radius $r$ is calculated
\be
E_M(r) = \frac{\pi}{\kappa} (4\Lambda r^2 + 4 Q^2  - 2 \mu \frac{Q^4}{r^2})
\,\,\label{Energy.M} \, . \en

Just as we expected, if $\mu \rightarrow 0$ the results expressed in (\ref{Energy}) and (\ref{Energy.M}) are in agreement with the energy distributions in both prescriptions for the 2+1 dimensional charged black hole as was evaluated in Ref.~\cite{Yang:2003vh}.

\vspace{.4cm}
\section{Energy distribution for another 2+1 dimensional static black hole solutions coupled to nonlinear electric field}

In the following paragraphs we consider two different 2+1 dimensional static black-hole solutions, both
coupled to nonlinear electric field and where the metrics have the form of Eq.~(\ref{metric}). As in the above case, we calculate the corresponding distributions of energy under both considered prescriptions.

Let us consider the solution 2+1 dimensional black hole coupled to the
Born-Infeld electrodynamics~\cite{cat.gar}
\be
A(r) = -M - (\Lambda - b^2)r^2 - 2 b^2 r \sqrt{r^2+\frac{Q^2}{b^2}} -
2 Q^2 \ln (r + \sqrt{r^2+\frac{Q^2}{b^2}})
\,\,\label{cat.gar} \, , \en
which can be rewritten as
\be
A(r) = -M - (\Lambda - b^2)r^2 + 4 b^2 \int^s_r \sqrt{u^2 + \frac{Q^2}{b^2}} \, du -C_1
\,\,\label{cat.gar-b} \, , \en
where $\Lambda$ is the cosmological constant, $b$ is the Born-Infeld parameter, $s$ is a constant and $C_1 = 2 b^2 s \sqrt{s^2+\frac{Q^2}{b^2}} +
2 Q^2 \ln (s + \sqrt{s^2+\frac{Q^2}{b^2}})$. Then
\be
E_E(r) = \frac{\pi}{\kappa}\left[-M - (\Lambda - b^2)r^2 + 4 b^2 \int^s_r \sqrt{u^2 + \frac{Q^2}{b^2}} \, du
- C_1 -1\right]
\,\,\label{E.cat.gar} ,  \en
\be
E_M(r) = \frac{\pi}{\kappa}\left[4(\Lambda - b^2)r^2 + 8 b^2 r \sqrt{r^2 + \frac{Q^2}{b^2}}\right]
\,\,\label{E.M.cat.gar} \, . \en

The second considered solution~\cite{cat.gar2} is given by
\be
A(r) = -M - \Lambda r^2 - Q^2 \ln(r^2 + a^2)
\,\,\label{cat.gar2} \, , \en
where $M$, $a$, $Q$ and $\Lambda$ are free parameters.
Rewriting, we have
\be
A(r) = -M - \Lambda r^2 + Q^2 \int_r^s \frac{2u}{(u^2 + a^2)}du - Q^2 \ln(s^2 + a^2)
\,\,\label{cat.gar2-b} \, . \en
The energy distributions become
\be
E_E(r) = \frac{\pi}{\kappa}[-M -\Lambda r^2 + Q^2 \int_r^s \frac{2u}{(u^2 + a^2)}du - Q^2 \ln(s^2+a^2)-1]
\,\,\label{E.cat.gar2} \,  \en
\be
E_M(r) = \frac{\pi}{\kappa}\left[4\Lambda r^2 + \frac{4Q^2r^2}{(r^2 + a^2)}\right]
\,\,\label{E.M.cat.gar2} \, . \en

The above results suggest we can establish for the $2+1$~case, similar to Ref.~\cite{Vagenas:2006pj} for $3+1$ dimensional static and spherically symmetric black holes (also see Ref.~\cite{Matyjasek:2008ku}), a connection between the coefficients $\alpha_n^{(E)}$ of the energy distribution under the Einstein prescription
\be
E_E(r) = \sum^{\infty}_{n=-2}\alpha_n^{(E)} r^{-n}
\,\,\label{sum-E} \,  \en
and the coefficients $\alpha_n^{(M)}$ of the energy distribution under the M{\o}ller prescription
\be
E_M(r) = \sum^{\infty}_{n=-2}\alpha_n^{(M)} r^{-n}
\,\,\label{sum-M} \, . \en
That relation is given by
\be
\alpha_n^{(M)} = 2n\alpha_n^{(E)}
\,\,\label{relation} \, . \en
If the the energy distribution has any term proportional to $\int F(u)\, du - C$, it can be written as
\be
\alpha_0^{(E)} = \int_r^s F(u) \, du  - C
\,\,\label{aE} \,  \en
and therefore
\be
\alpha_0^{(M)} = 2 \, r \, F(r)
\,\,\label{aM} \, . \en

In what follows we establish the relation between Eqs.~(\ref{Energy}) and (\ref{Energy.M}) having
in mind the above relations. To establish it we rewrite the Eq.~(\ref{fn}) as
\be
A(r) = -M -\Lambda r^2 + 2 Q^2 \int^s_r \frac{1}{u} \, du - 2 Q^2 \ln(s)
- \frac{\mu Q^4}{2 r^2}
\,\,\label{fnB} . \en
Considering Eq.~(\ref{fnB}), the energy distribution in Einstein prescription is
\be
E_E(r) = \frac{\pi}{\kappa}[-M-\Lambda r^2 + 2 Q^2 \int^s_r \frac{1}{u} \, du
-2 Q^2 \ln(s) - \mu \frac{Q^4}{2 r^2} -1]
\,\,\label{EnergyB} \, . \en
Note that this expression is equivalent to Eq.~(\ref{Energy}). Now if we use the Eqs.~(\ref{relation}-\ref{aM}) we get the following energy distribution
\be
E_M(r) = \frac{\pi}{\kappa}[4\Lambda r^2 + 4 Q^2 - 2 \mu \frac{Q^4}{r^2}]
\,\,\label{Energy.MB} \, , \en
which is similar to (\ref{Energy.M}) obtained under the M{\o}ller prescription. Additionally, the relation given by (\ref{relation}) is valid for the other considered cases in Ref.~\cite{Yang:2003vh}. One may also check the relation for other 2+1 dimensional black hole solutions with metric of the form (\ref{metric}). For example, the cases considered in Refs.~\cite{cataldo} and \cite{AyonBeato:2001sb} show that the relation (\ref{relation}) is fulfilled.

\vspace{.4cm}
\section{Summary}

In the previous sections we have obtained energy distributions for several 2+1 dimensional static black-hole solutions coupled to nonlinear electric field. First, we have presented the metric function (\ref{fn}) obtained from the first contribution of the weak field limit of one loop QED in 2+1 dimensions. We have calculated the energy distributions associated with the obtained metric using Einstein and M{\o}ller prescriptions. Similar calculations have been accomplished for another static black hole solutions in 2+1 dimensions coupled with nonlinear electrodynamics~\cite{cat.gar,cat.gar2}.
We have established when the energy distributions are of form given by expressions~(\ref{sum-E}) and
(\ref{sum-M}) there is a relation that connects the coefficients $\alpha_r$ of the Einstein prescription with those
in the M{\o}ller prescription. In summary, we showed that if the metric has the form of Eq.~(\ref{metric}), the energy distributions in the Einstein and M{\o}ller prescriptions are, respectively
\be
E_E(r) = \frac{\pi}{\kappa}[A(r) -1]
\,\,\label{EnergyE.gral} \, ,  \en
\be
E_M(r) = -\frac{2\pi}{\kappa} \, r A'(r)
\,\,\label{EnergyM.gral} \, . \en
Therefore, if we consider the expressions (\ref{sum-E}) and (\ref{EnergyE.gral}) we obtain
\be
A'(r) = \sum^{\infty}_{n=-2}\frac{\kappa}{\pi}\left[(\alpha_{n}^{(E)})' \,  r^{-n}- n \, \alpha_n^{(E)} r^{-(n+1)}\right]
\,\,\label{der.coeff} \, .  \en
Thus, substituting (\ref{der.coeff}) in (\ref{EnergyM.gral}) and considering the form (\ref{sum-M}) for the M{\o}ller energy distribution, we arrive to
\be
\alpha_n^{(M)} =  2 \, n \, \alpha_n^{(E)}
\,\,\label{coeff.gral} \,   \en
and
\be
\alpha_0^{(M)} =  -2 \, r\, (\alpha_{0}^{(E)})'
\,\,\label{coeff.gral-b} \,  , \en
when $\alpha_{0}^{(E)} = \int_r^s F(u) du$.

\vspace{.4cm}
\section*{Acknowledgments}

The author wish to thank Ver\'onica Raspa for her useful suggestions on the manuscript. This research was supported by a postdoctoral fellowship of
$\mbox{CONICET}$.


\vspace{.4cm}
\begin{thebibliography}{0}

\bibitem{einstein}
A. Einstein, Preuss. Akad. Wiss. Berlin {\bf 47}, 778 (1915).

\bibitem{tolman}
R. C. Tolman, \textit{Relativity, Thermodynamics and Cosmology}, Oxford Univ. Press (1934).

\bibitem{papapetrou}
A. Papapetrou,
Proc. R. Ir. Acad. {\bf A 52}, 11 (1948).

\bibitem{landau}
L. D. Landau and E. M. Lifshitz, \textit{The Classical Theory of Fields}, Addison-Wesley Press, Reading MA, 317 (1951).

\bibitem{bergman}
P. G. Bergmann and R. Thomson, Phys. Rev. {\bf 89}, 400 (1953).

\bibitem{weinberg}
S. Weinberg, \textit{Gravitation and Cosmology: Principles and aplications of General Theory of Relativity}, Wiley, New York, 165 (1972).

\bibitem{misner}
C. W. Misner, K. Thorne and J. A. Wheeler, \textit{Gravitation}, Freeman, San Francisco (1973).

\bibitem{qadir}
A. Qadir and M. Sharif, Phys. Lett. {\bf A167}, 331 (1992).

\bibitem{moller}
  C. M{\o}ller,
  Ann. Phys. (NY) {\bf 4}, 347 (1958).

\bibitem{Vagenas:2006pj}
  E.~C.~Vagenas,
  Mod.\ Phys.\ Lett.\  A {\bf 21}, 1947 (2006).

\bibitem{Matyjasek:2008ku}
  J.~Matyjasek,
  Mod.\ Phys.\ Lett.\  A {\bf 23}, 591 (2008).

\bibitem{virbhadra95}
K. S. Virbhadra, Pramana
{\bf 44}, 317 (1995).

\bibitem{Yang:2003vh}
  I.~C.~Yang and I.~Radinschi,
  AIP Conf.\ Proc.\  {\bf 895}, 325 (2007).

\bibitem{Radinschi:2007tt}
  I.~Radinschi and I.~C.~Yang,
  arXiv:gr-qc/0702105.

\bibitem{Grammenos:2004ys}
  T.~Grammenos,
  Mod.\ Phys.\ Lett.\  A {\bf 20}, 1741 (2005).

\bibitem{Radinschi:2005cj}
  I.~Radinschi and T.~Grammenos,
  Int.\ J.\ Mod.\ Phys.\  A {\bf 21}, 2853 (2006).

\bibitem{Vagenas:2003fc}
  E.~C.~Vagenas,
  Int.\ J.\ Mod.\ Phys.\  A {\bf 18}, 5949 (2003).

\bibitem{Vagenas:2004zt}
  E.~C.~Vagenas,
  Int.\ J.\ Mod.\ Phys.\  D {\bf 14}, 573 (2005).

\bibitem{brown}
J. D. Brown, J. Creighton and R. B. Mann, Phys. Rev. D{\bf 50}, 6394 (1994).

\bibitem{Balasubramanian:1999re}
  V.~Balasubramanian and P.~Kraus,
  Commun.\ Math.\ Phys.\  {\bf 208}, 413 (1999).

\bibitem{Miskovic:2006tm}
  O.~Miskovic and R.~Olea,
  Phys.\ Lett.\  B {\bf 640}, 101 (2006).

\bibitem{Sousa:2003sx}
  A.~A.~Sousa and J.~W.~Maluf,
  Prog.\ Theor.\ Phys.\  {\bf 108}, 457 (2002).

\bibitem{btz}
M. Ba\~nados, C. Teitelboim and J. Zanelli, Phys. Rev. Lett.
{\bf 69}, 1849 (1992).

\bibitem{cat.gar}
M. Cataldo and A. Garc\'{\i}a, Phys. Lett.
B {\bf 456}, 28 (1999).

\bibitem{cat.gar2}
M. Cataldo and A. Garc\'{\i}a, Phys. Rev.
D {\bf 61}, 084003 (2000).

\bibitem{Vagenas:2003ug}
  E.~C.~Vagenas,
  Int.\ J.\ Mod.\ Phys.\  A {\bf 18}, 5781 (2003).

\bibitem{Ritz:1995nt}
  A.~Ritz and R.~Delbourgo,
  Int.\ J.\ Mod.\ Phys.\  A {\bf 11}, 253 (1996).

\bibitem{DeLorenci:2001bd}
  V.~A.~De Lorenci, N.~Figueiredo, H.~H.~Fliche and M.~Novello,
  Astron.\ Astrophys.\  {\bf 369}, 690 (2001).

\bibitem{vonfreud}
P. von Freud, Ann. Math.
{\bf 40}, 417 (1939).

\bibitem{cataldo}
M. Cataldo, N. Cruz, S. del Campo and A. Garc\'{\i}a, Phys. Lett.
B {\bf 484}, 154 (2000).

\bibitem{AyonBeato:2001sb}
  E.~Ayon-Beato, A.~Garcia, A.~Macias and J.~M.~Perez-Sanchez,
  Phys.\ Lett.\  B {\bf 495}, 164 (2000).

\end{thebibliography}
\end{document}